\documentclass[12pt]{article}
\usepackage[tbtags]{amsmath}
\usepackage{amssymb}
\usepackage{amsthm}
\usepackage{array}
\usepackage{xy}
\usepackage{slashed}

\usepackage[iso-8859-7]{inputenc}
\usepackage{graphicx}

\def\<{\left\langle}
\def\>{\right\rangle}

\def\s{\sigma}

\def\p{\partial}

\newcommand{\be}{\begin{eqnarray}}
\newcommand{\ee}{\end{eqnarray}}
\newcommand{\bi}{\begin{itemize}}
\newcommand{\ei}{\end{itemize}}

\title{ {\ } \\[0.6cm] \Large
\bfseries Higher Derivative D-term Inflation \\[0.3cm] in New-minimal Supergravity  
\normalfont \\[1cm]
{\large  Iannis Dalianis$^{a}$ and Fotis Farakos$^{b}$} \\[0.6cm]
{\normalsize {\it  $^{a}$ Physics Division, National Technical University of Athens, \\[-0.2cm]15780 Zografou Campus, Athens, Greece}}\\[.4cm]
{\normalsize { \it $^{b}$ Institute for Theoretical Physics, Masaryk University, \\[-0.2cm] 611 37 Brno, Czech Republic}}\\[.2cm]
{\normalsize { E-mail: dalianis@mail.ntua.gr, fotisf@mail.muni.cz }}
}

\begin{document}

\date{}

\newcommand{\g}{\greektext} 
\newcommand{\e}{\latintext}
\maketitle

{\ } \\[-2cm]

\abstract{ 
We revisit the D-term inflation and amend it
with ghost-free higher derivative couplings of chiral superfields to super-curvature.  
These couplings realize a more generic inflationary phase in supergravity. 
After pointing out that a consistent embedding of these specific higher derivatives is known to exist only 
in the new-minimal supergravity, 
we show how a potential for the scalar component may arise due to a Fayet-Iliopoulos D-term. 
We then turn to inflationary cosmology and explicitly discuss different types of potentials, 
which 
capture properties of the common scenarios.  
These models  thanks to the 
derivative coupling:  i) naturally evade 
the supergravity $\eta$-problem, 
ii) drive inflation for a wider range of parameter values, and iii) predict lower values for the tensor-to-scalar ratio.}



\section{Introduction}
 
Observational data strongly indicate that an inflationary phase did occur at some stage in the early universe. 
Either in supersymmetric or non-supersymmetric theories the slow-roll
 inflation is the dominant paradigm \cite{Liddle:2000cg, Mukhanov:2005sc, Lyth:1998xn}. 
This phase is characterized by the Hubble friction, 
hence theories that generate enhanced friction effects are cosmologically
 rather motivated. 
It has been found that when a scalar field has derivative couplings to curvature, 
then it can slow-roll down even at relatively steep potentials during a
 (nearly) de Sitter phase. 
Nevertheless, not all derivative couplings to curvature are consistent, 
but there exist specific classes which lead to viable field theories \cite{Horndeski:1974wa}. 
An example of these ghost-free higher derivatives is the kinetic coupling of a scalar field to 
the Einstein tensor
 \begin{equation}
\label{GFF}
\frac{1}{M_*^2} \, G^{m n} \, \p_m \phi \, \p_n \phi \, 
\end{equation}
which has given rise to a considerable amount of scientific activity \cite{Amendola:1993uh,Capozziello:1999xt,Sushkov:2009hk,
Germani:2010hd,Germani:2010gm,Germani:2010ux,Saridakis:2010mf,Germani:2011ua,Charmousis:2011bf, Charmousis:2011ea,Koutsoumbas:2013boa,Kolyvaris:2011fk,Dent:2013awa}.  
In fact there also exist even higher order consistent derivative couplings \cite{Horndeski:1974wa,Nicolis:2008in,Deffayet:2009wt,Deffayet:2009mn,Deffayet:2010zh}.

The coupling (\ref{GFF}) has been called ``Gravitationally Enhanced Friction"
 (GEF) mechanism \cite{Germani:2011ua}. 
The attraction of the  mechanism is that it can set more general initial
 conditions for the inflationary phase, 
by relaxing the slow-roll conditions. 
Note that the mass scale in (\ref{GFF}) has to be smaller than the Hubble
 scale during inflation because only if $M_* \ll H$
the enhanced friction effects are more influential and noticeable. 
This raises the question of the origin of this scale; 
it is rather motivating to find it among the  dilaton couplings of the heterotic string \cite{Gross:1986mw}.

Thus, it would be desirable to have an {\itshape embedding} of the GEF mechanism 
in the theory of supergravity,
which is the framework under which supersymmetric theories during inflation  should be
 studied. 
In a supergravity theory, which is a non-renormalizable theory, the number of
 couplings that need to be specified is in principle infinite. The cut-off of
 the theory is considered to be the Planck mass, $M_P$,  and an estimation in
 supergravity ceases to be reliable for field values $\phi \gg M_P$, 
 unless the non-renormalizable terms are suppressed, e.g. due to a particular symmetry.
Even though one can construct models in which the inflaton field value
 experiences sub-Planckian variation, 
 a generic supergravity theory will fail to inflate 
because the F-term part of the potential yields a too big inflaton mass \cite{Lyth:1998xn}.

The embedding of the GEF mechanism  is in general not a trivial task; 
as we have mentioned, higher derivative theories 
may come accompanied by ghost instabilities. 
In earlier works \cite{Farakos:2012je,Farakos:2013fne} it was understood, 
that in order for this coupling to be consistently realized, 
one has to turn to the new-minimal supergravity  \cite{Sohnius:1981tp,Sohnius:1982fw,Ferrara:1988qxa}. 
Nevertheless, the coupling is still inconsistent unless the non-minimally
 coupled superfield 
has a vanishing R-charge, and is neutral under any gauge group. 
Hence it is not possible to endow this superfield with a conventional self-interaction in terms of superpotential or gauging. 
This problem can be solved, as we propose here, by breaking supersymmetry with a Fayet-Iliopoulos
 term, 
which may induce a potential for the scalar field.

In this work we explore the modified dynamics of a non-minimally coupled
 superfield to curvature,
 and we find that inflation can be realized and described  reliably in a
 supergravitational framework. 
Indeed, the scalar component of the $\Phi$ superfield is governed solely by a
 D-term potential and 
experiences a high friction during a de Sitter phase. 
Moreover, due the vanishing of the superpotential, it is also expected to have 
 negligible interactions with other fields, 
a fact that is supported by the Planck observational data.
Therefore the superfield $\Phi$ is tailor-made for driving  inflation in
 supergravity.

\vskip0.2cm
 
The motivation 
of this article is both particle theoretical and cosmological.
On technical grounds, in section 2, we show how it is possible to introduce a potential for
 the non-minimally coupled field when it
is coupled to supergravity. Then, in section 3, we show how an inflating theory driven by the supersymmetric
 slotheon - it has been named slotheon after \cite{Germani:2011bc}, 
can evade some shortcomings common in conventional inflationary  supergravity and we revisit particular inflationary
 models. We find the new field space region where an accelerated expansion is
 realized and check whether these models can fit the observational data even
 though they were previously excluded.

\section{New-minimal supergravity: Derivative couplings and D-terms }

The minimal theories of supergravity have a rich structure 
originating from the  possible compensating multiplets that break the
 underlying 
superconformal theory to super-Poincar\'e \cite{Gates:1983nr,Buchbinder:1998qv}. 
The underlying dualities among the compensating multiplets 
survive the gauge fixing and lead to equivalent couplings to matter \cite{Ferrara:1983dh}, 
but break down as soon as higher derivatives are introduced. 
The couplings we want to study here make this duality-breakdown even more
 manifest, 
since the only known supergravity which can accommodate them 
in a consistent way \cite{Farakos:2012je,Farakos:2013fne}
is the so-called {\it new-minimal supergravity} \cite
 {Sohnius:1981tp,Sohnius:1982fw,Ferrara:1988qxa}.
An aspect of the new-minimal supergravity not encountered in the 
{\it standard } supergravity is the necessary existence of a chiral symmetry. 
It is well known that rigid supersymmetry allows for the existence of a
 {\itshape chiral symmetry} called $U(1)_\text{R}$. 
This {\itshape R-symmetry} becomes local 
and is gauged by
 one of the auxiliary fields of the gravitational supermultiplet.

The  new-minimal supergravity  \cite{Sohnius:1981tp} is the supersymmetric theory of the gravitational multiplet
\begin{equation}
e^a_m\, ,~~\psi_m^\alpha \,  , ~~A_m\, , ~~ B_{mn}\ .
\end{equation}
The first two fields are the vierbein and its superpartner the gravitino, 
a spin-$\frac{3}{2}$ Rarita-Schwinger field. 
The last two fields are auxiliaries. 
The real auxiliary vector $A_m$ gauges the $U(1)_\text{R}$ chiral symmetry. 
The auxiliary $B_{mn}$ is a real two-form appearing only through its dual field strength $H_m$, 
which satisfies $ \hat {\cal D}^a H_a =0$,
for the supercovariant derivative $\hat {\cal D}^a$.  
This constraint can be solved in terms of $B_{mn}$.
Note that all the fields of the new-minimal supergravity multiplet are gauge fields.

We will use superspace techniques to guarantee that our component form 
Lagrangians are supersymmetric.  
The interested reader may consult for example \cite{Ferrara:1988qxa} where a treatment of 
the new-minimal superspace is given. 
The new minimal supergravity free Lagrangian is given by
\begin{equation}
\label{sugra} 
{\cal{L}}_{\text{sugra}}= - 2 M^2_P  \int d^4 \theta E V_{\text{R}} . 
\end{equation}
Here $V_{\text{R}}$ is the gauge multiplet of the R-symmetry, 
which (in the appropriate WZ gauge) contains the auxiliary fields in its vector component, $- \frac{1}{2} [\nabla_\alpha , \bar \nabla_{\dot \alpha} ]  V_{\text{R}} | = A_{\alpha \dot \alpha} - 3 H_{\alpha \dot \alpha}$, 
and the Ricci scalar in its highest components, $\frac{1}{8} \nabla^\alpha \bar \nabla^2 \nabla_\alpha  V_{\text{R}} |= - \frac{1}{2} \left( R + 6 H^a H_a \right)$. 
The $E$ is the super-determinant of new-minimal supergravity, 
but in general (as we also do here) 
one can calculate the supersymmetric Lagrangians only with the use of 
the F-term formula, since
\be
 \int d^4 \theta E X = \frac{1}{2} \int d^2 \theta {\cal E} \left( -\frac{1}{4} \bar \nabla^2 X \right) + c.c.
\ee
In the chiral {\it theta} expansion the chiral density reads ${\cal E} = e  + i e \sqrt 2 \theta \s^a \bar \psi_a - \theta^2 e \bar \psi_a \bar \s^{ab} \bar \psi_b$.
Note that $X$ is a generic hermitian superfield with vanishing chiral weight, 
while its chiral projection ($ -\frac{1}{4} \bar \nabla^2 X$) has chiral weight $n=1$. 
The bosonic sector of Lagrangian (\ref{sugra}) is
\begin{equation}
\label{freesugra}
{\cal{L}}_{\text{sugra}}^B= 
  M^2_P \, e \, \left( \frac{1}{2}
R + 2A_a H^a-3H_a H^a\right) .
\end{equation}

For the matter sector we have a chiral multiplet, defined by
$\bar \nabla_{\dot \alpha} \Phi =0$ 
which has bosonic components, a physical complex scalar $A=\phi+i\beta$, 
and an auxiliary complex field $F$, defined as  
\begin{equation}
\Phi | = A
\  ,\ \quad
-\frac{1}{4} \nabla^2 \Phi | = F .
\end{equation}
In general, a chiral superfield in new-minimal supergravity is allowed to have an arbitrary R-charge,  
but we stress that our chiral superfield has a vanishing one  \cite{Farakos:2012je} in order to avoid ghost instabilities
\begin{equation}
n_\Phi =0.
\end{equation} 
 The minimal kinematic Lagrangian for this multiplet is in superspace 
\begin{equation}
{\cal L}_0 = \int d^4  \theta \, E  \bar \Phi \Phi 
\end{equation}
the bosonic sector of which is
\begin{equation}
{\cal L}^B_0 = A\Box \bar A+  F\bar F 
-i H^m \left(A\,\partial_m \bar A- \bar A \partial_m A\right) .
\end{equation}
Finally, concerning our chiral superfield, it will also have a non-minimal derivative coupling with the supergravity 
multiplet
\begin{equation}
\label{NMDC}
{\cal L}_{M_*} =  i M^{-2}_* \int d^4 \theta \, E \, [ \bar \Phi E^a \nabla_a \Phi ] + c.c.  
\end{equation}
where $E^a$ is a curvature real linear superfield ($\nabla^2 E_a = \bar \nabla^2 E_a =0$) of the new-minimal supergravity. 
The $E^a$ superfield has bosonic components  $E_a | = H_a$
and 
$\frac{1}{4}  \bar \s_a^{\dot \alpha \alpha}  [\nabla_\alpha , \bar \nabla_{\dot \alpha} ] E_b | = \frac{1}{2} ( G_{ab} - g_{ab} H^c H_c - 2 H_a H_b - {}^* {\cal F}_{ab} )$, 
where $G_{mn} = R_{mn} - \frac{1}{2} g_{mn} R$ is the Einstein tensor and 
$ {\cal F}_{mn} = \p_m A_n - \p_n A_m $ is the field strength of the supergravity auxiliary field $A_m$. 
It is worth mentioning that $E^a$ satisfies the superspace Bianchi identity 
$\nabla^a E_a =0$.
For a discussion and derivation of the Lagrangian (\ref{NMDC}) see \cite{Farakos:2012je}. 
The bosonic sector of  Lagrangian (\ref{NMDC}) is 
\begin{equation}
\begin{split} 
{\cal L}^B_{M_*} = & M^{-2}_* \left[ 
\, G^{ab}\partial_b \bar A \,\partial_a A
+2F \bar F H^a A_a
-2F \bar F H^aH_a\right.
\\
\label{non-min}
&
+i  H^a \left(\bar F \partial_a F-F\partial_a \bar F\right) 
- \partial_bA\,\partial^b \bar A H_aH^a \\
& \left. +2\,H^a\partial_aA\,H^b\partial_b \bar A
-i H_c \left(\partial_b \bar A \,{\cal{D}}^c\partial^b A 
-\partial_b A\,{\cal{D}}^c\partial^b \bar A \right)\right] . 
\end{split}
\end{equation}
Note that this term, although it contains higher derivatives, 
does not lead to ghost states or instabilities. 
The ghost instabilities are in fact evaded due to the vanishing chiral weight of the chiral superfield $\Phi$.  
On the other hand, 
the vanishing chiral weight forbids the self-coupling via a superpotential due to the R-symmetry. 
Thus this superfield is not allowed to have a superpotential. 
Moreover it is also not allowed to be gauged, 
since this will also give rise to ghost instabilities 
via inconsistent derivative couplings of the gauge fields to curvature. 
The only remaining option 
is the indirect introduction of self-interaction via a  {\it gauge kinetic function}.

The gauge sector of our theory is composed of a standard $U(1)$ gauge multiplet $V$,  
with a $\Phi$-dependent gauge kinetic function and a Fayet-Iliopoulos term. 
The $U(1)$ gauge multiplet 
consists of a gauge vector field $v_m$, a majorana gaugino $\lambda^{\alpha}$, 
and a real auxiliary field $D$. 
In particular, the definition of the bosonic components of the vector is
\be
- \frac{1}{2} [\nabla_\alpha , \bar \nabla_{\dot \alpha} ]  V | = v_{\alpha \dot \alpha} \ , \ 
 \frac{1}{8} \nabla^\alpha \bar \nabla^2 \nabla_\alpha  V |= D . 
\ee
Note that due to the structure of new-minimal supergravity, a FI term is in general allowed 
(even the superspace Lagrangian of pure  new-minimal supergravity (\ref{sugra}),  {\it is} a FI term). 
Thus we have in superspace
\begin{equation}
\label{GGG}
{\cal L}_g = \frac{1}{4} \int d^2 \theta {\cal E} f(\Phi) W^2 (V) + c.c. + 2 \xi \int d^4 \theta E V
\end{equation}
with $W_{\alpha} (V) = - \frac{1}{4} \bar \nabla^2 \nabla_\alpha V $.
The $f(\Phi)$ is a holomorphic function of the chiral superfield $\Phi$ 
and $\xi$ is the Fayet-Iliopoulos parameter of mass dimension two. 
The bosonic sector of (\ref{GGG}) reads
\begin{equation}
\label{g}
e^{-1} {\cal L}_{\text{g}}^B= 
- \frac{1}{4}  \text{Re}f(A) F^{mn} F_{mn} 
+ \frac{1}{4} \text{Im}f(A) F^{mn} \ ^*F_{mn}
+\frac{1}{2} \text{Re}f(A) D^2
+ \xi D 
-2 \xi  v_a H^a 
\end{equation}
where $F_{mn} = \p_m v_n - \p_n v_m$.
We stress that the scalar $A$ is not charged under this $U(1)$, 
thus there is no restriction in the form of $f(A)$ apart from holomorphicity.

The total Lagrangian we are interested in is 
\begin{equation}
\label{total}
{\cal{L}}_{\text{total}}
=M^2_P \, {\cal{L}}_{\text{sugra}}
+ {\cal L}_{\text{g}}
+ {\cal L}_0
+ {\cal L}_{M_*}
\end{equation}
which reads

\begin{equation}
\label{total'}
\begin{split}
e^{-1}{\cal{L}}_{\text{total}} 
& = M^2_P
\left[ \frac{1}{2}{{\cal{R}}}+2
V^a\,H_a-3H^aH_a\right] + A\Box
\bar A+     F \bar F  \\ 
&+M_*^{-2}\left[
\, G^{ab}\partial_b
\bar A \,\partial_a A-2F \bar F H^aH_a 
- \partial_bA\, \partial^b \bar A \,     
H_aH^a +2H^a\partial_aA\,H^b\partial_b
\bar A\right]\\
& -\frac{1}{4}  \text{Re}f(A) F^{mn} F_{mn} 
+ \frac{1}{4} \text{Im}f(A) F^{mn} \ ^*F_{mn}
\\
& +\frac{1}{2} \text{Re}f(A) D^2
+ \xi D\,,
\end{split}
\end{equation}
where we have redefined the auxiliary field $A^a$ to $V^a$ 
\begin{equation}
\label{V0}
\begin{split}
V^a= & A^a\Big{(}1+\frac{1}{M^2_P} M^{-2}_* F\bar F\Big{)} - \frac{1}{M^2_P} \xi v^a
\\
&+ \frac{1}{2M^2_P}\Big{(}i \bar A
\partial^aA-iA
\partial^a \bar A- iM^{-2}_*F
\partial^a\bar F+iM^{-2}_*\bar F
\partial^aF \Big{)}
\\
& 
 + \frac{1}{2M^2_P}\Big{(}  iM^{-2}_*\partial_bA\,{\cal{D}}^a\partial^b \bar A 
-iM^{-2}_*\partial_b\bar A{\cal{D}}^a\partial^b A\Big{)}.
\end{split}
\end{equation}

Lagrangian (\ref{total'}) contains four auxiliary fields.
First,  by solving the equations of motion for the supergravity auxiliary fields
we find that the vector $H_m$ vanishes and $V_m$ reduces to a pure gauge. 
In fact since $H_m$ is the dual field-strength of $B_{mn}$ here both auxiliary fields of the
new-minimal supergravity are pure gauge on-shell. 
For the auxiliary field $F$ it is easy to see that it will also vanish on-shell.
Finally, by solving the equations of motion for the auxiliary 
field $D$ of the gauge multiplet we find
\begin{equation}
\label{D}
D= - \frac{\xi}{\text{Re}f(A)}.
\end{equation}
After plugging back our results, we have the following on-shell form for (\ref{total'})
\begin{equation}
\label{total-onshell}
\begin{split}
e^{-1}{\cal{L}}_{\text{total}} =& \frac{M^2_P}{2} {{\cal{R}}}  + A\Box
\bar A 
+M^{-2}_* 
\, G^{ab}\,\partial_a
\bar A\, \partial_b A\ -\frac{1}{2} \frac{\xi^2}{\text{Re}f(A)} 
\\
& -\frac{1}{4}  \text{Re}f(A) F^{mn} F_{mn} 
+ \frac{1}{4} \text{Im}f(A) F^{mn} \ ^*F_{mn}.
\end{split}
\end{equation}
Note that this Lagrangian (\ref{total-onshell}) does {\it not} contain ghost states or instabilities.

The Fayet-Iliopoulos term inside (\ref{g}) breaks supersymmetry, 
and combining it with the gauge kinetic function 
has the effect of introducing a scalar potential which reads
\begin{equation}
\label{VA}
{\cal V} = \frac{1}{2} \frac{\xi^2}{\text{Re}f(A)},
\end{equation}
where $ A=\phi+i\beta$. This is the D-term potential. 
It is expected that only in the case of broken supersymmetry one 
can have a potential for the  $A$ field which has the non-minimal kinetic coupling to gravity.
This stems from the fact that the chiral $U(1)_\text{R}$ symmetry 
of new-minimal supergravity forbids a potential for this field, 
due to its vanishing chiral weight.
The advantage of a Fayet-Iliopoulos term is that it breaks supersymmetry spontaneously.

\section{Application to inflation}

\subsection{A pure D-term inflation}

In the standard supergravity the scalar potential of chiral superfields transforming in some representation of a gauge group has the following form\footnote{This formula is common for the old-minimal supergravity \cite{Wess:1992cp}. 
Nevertheless, a general supergravity-matter system \\ in the new-minimal framework can be recast in this form after 
appropriate redefinitions \cite{Ferrara:1988qxa}.}
\begin{equation}
V=e^{K/M^2_P} \left[F_i (K^{-1})_j^i F^j -3 \frac{|W|^2}{M^2_P} \right] +\frac{g^2}{2} \frac{1}{\text{Re}f_{ab}} D^a D^b
\end{equation}
where $F^i=W^i+K^iW/M^2_P$ and $D^a=K^i(T^a)^j_i z_j + \xi^a$. The upper (lower) index $i$ denotes derivatives with respect to the $\phi_i$ ($\phi^{*i}$) field.  The slow-roll conditions imply 
\begin{equation}
\epsilon \ll 1 \Rightarrow \quad \frac{K_\phi}{M_P}+... \ll 1
\end{equation}
\begin{equation}
\eta \ll 1 \Rightarrow \quad 3K_{\phi \bar{\phi}} H^2+... \ll H^2\,.
\end{equation}
Here the subscript $\phi$ denotes a derivative with respect to the inflaton. The inflaton vacuum energy dominates the energy density of the universe and the relation $H^2=V/(3M^2_P)$ has been used in the second condition $\eta \ll 1$. In the low energy minimum the K\"ahler metric should be normalized to one and it is not expected to be suppressed during inflation. Therefore F-type inflation in supergravity theories is hard to be realized unless the K\"ahler and the superpotential have a special form or accidental cancellations take place \cite{Lyth:1998xn, Liddle:2000cg, Binetruy:1996xj}.
 
A resolution to this $\eta$-problem in generic supergravity theories can be given by a symmetry that suppresses the F-term part of the scalar potential. In the presence of such a symmetry the potential is naturally dominated by a Fayet-Iliopoulos D-term which exists for $U(1)$ gauge groups. Here, the R-symmetry of the theory forbids the superpotential interactions for the $A$ field non-minimally coupled to the $G^{mn}$ tensor. The spontaneous breaking of supersymmetry during inflation may introduce interactions however these will be generated radiatively and should not affect the tree level D-term inflationary potential. The D-term potential domination together with the enhanced friction features strongly motivates the study of this higher derivative theory to inflationary applications.

\subsection{Expanding the allowed initial conditions for inflation}

The complex scalar  field $A$ is governed by the scalar potential generated by the Fayet-Iliopoulos supersymmetry breaking and has the form (\ref{VA}).
In our context the gauge kinetic function $\text{Re}f(A)$ is arbitrary and in principle contains non-renormalizable terms. In most of the models, again, one finds that the $|A|$ is of order $M_P$ or larger, a fact that makes the non-renormalizable terms difficult to control similarly to the higher order terms in the $K$ and $W$ potential. Here, we will approximate $f(A)$ by  polynomials and monomials 
 or ascribe to it forms suggested by microscopic theories as the string theory. 
 
In a FLRW background, neglecting spatial gradients, the Friedmann equation and the equation of motion for the $\phi$ (or the $\beta$) field are
\begin{equation} \label{mod}
H^2=\frac{1}{3M^2_P}\left[\frac{\dot{\phi}^2}{2}\left(1+9M^{-2}_*H^2\right)+V(\phi) \right], \quad\quad \partial_t\left[a^3\dot{\phi}\left(1+3M^{-2}_*H^2 \right) \right]=-a^3 V_\phi\,.
\end{equation}
Let us first demonstrate the advantages of the kinetic coupling to the inflationary applications.  We assume here that the $\phi$ is the single inflating field and we consider a full polynomial potential without symmetry suppressed terms, actually non-renormalizable terms are naturally present in supergravity theories:
$V(\phi)=\sum_n\lambda_nM^{4-n}_P\phi^n$.
In the large field models of inflation the inflaton field has a value of order the Planck mass, $M_{P}$. This general potential cannot serve as large field inflationary model for the non-renormalizable terms, if not suppressed, spoil the flatness of the potential.

The non-minimal coupling of the kinetic term of the scalar field with the Einstein tensor $G_{mn}$ 
\begin{equation}
{\cal L}=-\frac{1}{2} \sqrt{-g} \left(g^{mn} -M^{-2}_*G^{mn} \right)\partial_m\phi\partial_n\phi
\end{equation}
{\itshape during a de Sitter phase} takes the simple form $M^{-2}_*G^{mn}=-3M^{-2}_*H^2 g^{mn}$. For $H M^{-1}_*\gg 1$ the kinetic coupling implies that the canonically normalized scalar field is the $\tilde{\phi}= \sqrt{3}H M^{-1}_* \phi$. 
This rescaling recasts the polynomial potential 
in terms of the canonically normalized inflaton $\tilde{\phi}$ to the form
\begin{equation}
V(\tilde{\phi})=\sum_n \lambda_n M^{4-n}_P \left(\frac{\tilde{\phi}}{\sqrt{3}HM^{-1}_*}\right)^n\,.
\end{equation}
The non-renormalizable terms $\sum^\infty_{n=4} \lambda_n M_{P}^{4} \left({\tilde{\phi}}\times({\sqrt{3}H M^{-1}_* M_{P}})^{-1}\right)^n $ are suppressed by the ``enhanced'' mass scale $\sqrt{3}H M^{-1}_* M_{P}$. The slow roll parameters require $\tilde{\phi}>M_{P}$ and, hence, these higher order terms can be neglected and sufficient inflation can take place given that 
\begin{equation} \label{fv1}
M_{P} < \tilde{\phi} \ll M_{P}(H M^{-1}_*)\,.
\end{equation}
In terms of the field $\phi$, which has non-canonical kinetic term, the above field-space region translates into
\begin{equation} \label{fv2}
\frac{M_{P}}{H M^{-1}_*} < {\phi} \ll M_{P}\,.
\end{equation}
This finding is of central importance since we work in a supersymmetric context. Even though we suggest a D-term inflation without a superpotential the generation of the inflationary potential is in principle not protected by any symmetry and in the most general case we cannot forbid the higher order terms. 

We consider our theory as an effective one valid below some ultra-violet cut-off that we generally identify with the $M_P$. The field-space region (\ref{fv2}) allows inflation to be realized in general form of potentials and reliable conclusions in this context can be derived. It can be said that the kinetic coupling theory is tailor-made for realizing an inflationary phase.

From a different perceptive, if there is an internal symmetry that forbids the non-renormalizable terms and thereby suppresses the coefficients $\lambda_n$ for $n\geq 5$ then inflation can be implemented in a much larger  field-space region, than in the conventional (GR limit) large field inflationary models, that reads:
$\phi>M_P/(HM^{-1}_*)$

\subsection{Introducing D-term inflationary potentials}

We will attempt to capture some of the characteristics of the kinetic coupling in inflationary applications by considering some representative examples of  inflationary potentials. We will concentrate on {\itshape single field inflation} models where one of the two fields is heavy enough and stabilized in the vacuum.  

According to the Eqs. (\ref{mod}) and for $H M^{-1}_*\gg 1$ the slow-roll parameters of General Relativity (GR) $\epsilon \equiv  {M^2_P} ({V'}/{V})^2/2$ and $\eta\equiv M^2_P V''/V$ are recast into 
\begin{equation}
\tilde{\epsilon} \approx \frac{\epsilon}{3H^2 M^{-2}_*} \,\, , \quad \quad \tilde{\eta} \approx \frac{\eta}{3H^2 M^{-2}_*}\,.
\end{equation}
The requirement $\tilde{\epsilon}, |\tilde{\eta}|<1$ yields that the field space region where slow-roll inflation is realized is rather increased. We will illustrate this below by considering different forms for the gauge kinetic function and thereby various types of potentials.

{\itshape Linear potentials.} 
Let us first assume that 
\begin{equation} \label{f1}
f(A) = \frac{\xi^2}{2V_0}\sum_n \lambda_n\left(\frac{ A}{M_P}\right)^n
\end{equation}
where $\lambda_n$ are real coefficients and we constrain the field to sub-Planckian values $|A| \ll M_P$. 
The scalar potential reads ${\cal V}(\phi, \beta) = V_0\left(1-\lambda_1{\phi}/{M_P}- (\lambda_2 -\lambda^2_1)\phi^2/M^2_P +\lambda_2\beta^2/{M^2_P}+...\right)$  where the ellipsis corresponds to negligible terms. The above potential includes two scalars that have a non-minimal derivative coupling. For $\lambda^2_1\sim \lambda_2>0$ the $\phi$ field can be light enough and the $\beta$ field can be heavy enough ($\gtrsim H$) and stabilized; hence the appearance of any sub-Planckian strong coupling scale \cite{Germani:2011ua} can be avoided. For $\phi\ll M_P$
 the linear to $\phi$ term dominates and the potential reads: 
\begin{equation} \label{hill}
{\cal V} \simeq V_0\left(1-\lambda \frac{\phi}{M_P}\right)\,.
\end{equation}
The slow-roll conditions 
 yield the requirements for inflation $\tilde{\eta}=0$ and $\tilde{\epsilon} \approx M^2_P\lambda^2/(2V_0 M_*^{-2} )<1$. 

{\itshape Exponential potentials.}
If we now assume that the gauge kinetic function is of exponential form then we directly get an exponential type potential:
\begin{equation} \label{fexp}
f(A)= \frac{\xi^2}{2V_0} e^{\lambda A/M_P} \quad \Rightarrow \quad {\cal V} =V_0 \frac{1}{\cos(\lambda \beta/M_P)} e^{-\lambda \phi/M_P}\,.
\end{equation}
The form of the function $1/\cos x$ suggests that the $\beta$-dependent part of the potential will be stabilized with large enough mass
to values  $\langle{1/\cos}(\lambda \beta/ M_P) \rangle = 1 $ and the $\phi$ will be the inflating field. An important reason for picking the gauge kinetic function (\ref{fexp}) is that it is reminiscent of the dilaton coupling of string theory. Slow-roll inflation takes place for $\tilde{\eta}=2\tilde{\epsilon}=\lambda^2/(3H^2 M_*^{-2}) <1$ which corresponds to inflaton field values $\phi< M_P/{\lambda} \ln \left({ V_0/ M_*^{2}}{M^2_P \lambda^2}\right)$, see Ref. \cite{Dalianis:2014nwa} for further analysis.

{\itshape Inverse power law potentials.}  
As a final example, an inverse power law potential can be obtained if we consider monomial gauge kinematic function:
\begin{equation} \label{finv}
f(A)= \frac{\xi^2}{2V_0}\, \lambda \frac{A^n}{M^n_P}\quad \Rightarrow \quad {\cal V} =V_0 \frac{1}{\lambda} \frac{M^n_P}{\phi^n}\,,
\end{equation}
where we used the relation Re$\{A^n\}=\phi^n \cos(n\theta)/ (\cos\theta)^n$. We see that the field $\theta$, the phase of the complex field $A=\rho e^{i\theta}$, is stabilized with large enough mass at $\theta =\kappa \pi$ and so Re$\{A^n\}=\phi^n$. 
Slow-roll inflation takes place for $\tilde{\epsilon}=(M^2_Pn^2)/(2\phi^2\times 3H^2 M_*^{-2})<1$, $\tilde{\eta}=2(1+n^{-1})\tilde{\epsilon}<1$ which corresponds to inflaton field values $\phi^{n-2}<2 M^{n-4}_P {V_0 }/{M_*^{2}\lambda n^2}$.
\\
\\
For the above types of potentials an inflationary phase can be realized for a wider range of parameters. For the linear and the exponential, in GR limit, inflation 
is impossible for $\lambda \geq {\cal O}(1)$. It has to be $\lambda<{\cal O}(1)$ which implies, after absorbing $\lambda$ to the mass scale, that the field $\phi$ has to be suppressed by a super-Planckian value. Here thanks to the kinetic coupling inflation is possible even for $\lambda \gtrsim 1$ and for sub-Planckian excursions for the (non-canonical) inflaton field $\phi$.

\subsection{Enhanced friction supergravity inflationary models and the CMB data}

In order to make contact between the theory and observation the spectra of scalar and tensor perturbations have to be estimated \cite{Liddle:2000cg, Langlois:2004de}.
The density perturbations $\delta \rho$ of the inflaton are encoded in the variable $\zeta=\delta\rho/(\rho + p)$ which is conserved on large scales in the absence of entropy perturbations and can be directly related to the cosmic microwave background temperature fluctuations. The power spectrum of the $\zeta$ variable in first order in the slow roll parameter $\tilde{\epsilon}$ reads  \cite{Germani:2011ua} ${\cal P}_\zeta \approx H^2/8\pi^2 \tilde{\epsilon} c_s M^2_P$.
The sound speed squared having a dependence $c^2_s \propto H^2 \tilde{\epsilon} $, is subluminal and modifies the spectral tilt dependence on the slow roll parameters. The result is
\begin{equation} \label{ns}
\left. n_s-1 \equiv \frac{d \ln {\cal P}_\zeta}{d \ln k}\right|_{c_sk=aH} \approx -8 \tilde{\epsilon} + 2\tilde{\eta}
\end{equation}
contrary to the well known GR limit formula $n_s-1=2\eta-6\epsilon$. The ratio of the tensor to scalar amplitudes, $r\equiv {\cal P}_g(k_*)/{\cal P}_\zeta(k_*)$, has the conventional GR dependence on the slow-roll parameter $\tilde{\epsilon}$ at the lowest order: $r=16 \tilde{\epsilon}$. However the new relation (\ref{ns}) allows for larger values for the $\tilde{\eta}$ slow-roll parameter. Namely it is 
\begin{equation}
r=2(1-n_s)+4 \tilde{\eta}
\end{equation}
instead of $r=(8/3)(1-n_s)+(16/3)\eta$. Hence, given that $1-n_s \cong 0.04$ and $r<0.11$ \cite{Ade:2013uln},  positive values for the $\tilde{\eta}$ can be accommodated, which correspond to potentials with $V'' \geq 0$.  The number of e-folds, ${N}\equiv\int Hdt$, for $H^2M^{-2}_* \gg 1$ is given by the expression 
\begin{equation}
 {N}(\phi)=\frac{1}{M^2_P} \int^{\phi}_{\phi_f} \left(1+3H^2M^{-2}_* \right)\frac{V}{V'} d\phi \approx \frac{1}{M^4_P} \int^{\phi}_{\phi_f} M^{-2}_*\frac{V^2}{V'} d\phi
\end{equation}

The Planck Collaboration estimated the spectral index $n_s$ from the observational data (Planck and WMAP) to be \cite{Ade:2013uln}
 $n_s=0.9603 \pm 0.0073$
 and the upper bound on the tensor to scalar ratio at $r<0.11$. This constraint on $r$ corresponds to an upper bound on the energy scale of inflation $H(\phi_*)/M_P \leq 3.7 \times 10^{-5}$ which implies that $\tilde{\epsilon}(\phi_*)\equiv \tilde{\epsilon}_*<0.008$. On the other hand, the BICEP2 Collaboration \cite{Ade:2014xna} reported a value $r\sim0.2$  which allows larger values for the $\tilde{\epsilon}_*$. The $\phi_*$ denotes the field value during inflation that the pivot scale $k_*=0.002 \text{Mpc}^{-1}$ \cite{Ade:2013uln} exited the Hubble radius (not to be confused with the subscript at the mass scale $M_*$ of the non-minimal derivative coupling).
\\
\\
Let us now examine the supergravity D-term inflationary models of the previous subsection in the light of the observational data. The key ingredient is that the inflaton field is characterized by the non-minimal derivative coupling to the Einstein tensor.

The {\itshape linear} model (\ref{hill}) 
yields a spectral index $1-n_s=8\tilde{\epsilon}=4\lambda^2M^2_P M^2_*/V_0$ which is related to the number of e-folds by the expression $1-n_s \approx 4\lambda \Delta\phi/N(\phi_*) M_P$. 
 For $N(\phi_*) \sim 50$ and $\Delta \phi \ll M_P$ this model can give a spectral index value $1-n_s \sim 0.04$ and $r\sim 0.08$ for $\lambda\gg 1$.  
For the {\itshape exponential potential} (\ref{fexp})
one finds $\tilde{\eta}=2\tilde{\epsilon}$, $1-n_s=8\tilde{\epsilon}-2\tilde{\eta}=4\tilde{\epsilon}$ and $1-n_s\sim 2/N(\phi_*)$. 
Hence, it predicts a tensor-to-scalar ratio $r=0.16$ for $1-n_s=0.04$. This value of $r$ lies between the Planck and the BICEP2 data.
In the case of the {\itshape inverse power-law} models (\ref{finv}) the spectral index is given by the modified  expression $1-n_s=4 \tilde{\epsilon}\, (1-n^{-1})$ and it is in tension with the Planck data however, it is viable according to the BICEP2 data for large enough power $n$.

The kinetic coupling operates like an enhanced friction and inflation takes place more generically than in the GR limit. Since inflation is primarily introduced to address (or better ameliorate) the homogeneity, isotropy and flatness problem  we can say that  even excluded models are still motivated candidates for inflation in the context of supergravity kinetic coupling. Afterwards, in order to seed the large-scale structure formation in the universe, a mechanism like the  curvaton or the modulated reheating may take place in the post-inflationary universe.

The non-minimal derivative coupling of the scalar field $\phi$  with the Einstein tensor in supergravity renders the $\phi$ a compelling inflaton candidate due to both the enhanced friction effect and the symmetry suppression of the F-terms. Despite these advantages the absence of tree level superpotential interactions may be problematic for a sufficient  reheating of the universe. However, we note that there is the coupling of the inflating scalar $\phi$ with the gauge field strength (\ref{total-onshell}) and, also, interactions beyond the  tree level.  Moreover, there are mechanisms like the gravitational particle production \cite{Ford:1986sy, Felder:1999pv} invented for such questionable situations.

\section{Conclusions}
In the present paper we have examined the implementation of an inflationary phase by the scalar component of a chiral superfield in a supergravity context. The particular characteristic of this scalar is that it is non-minimally coupled to the Einstein tensor $G^{mn}$,  and the slow-roll conditions can be satisfied more generically. The potential is introduced via a Fayet-Iliopoulos D-term since the R-symmetry of the theory excludes the introduction of a superpotential,  and its gauging is also forbidden due to stability issues.  Even though this is a higher derivative theory it does not give rise to ghost instabilities.  

The model we propose is a pure D-term inflation with a gravitational enhanced slow-roll for the inflaton. These two features have important implications for the inflationary dynamics. Firstly, the accelerated expansion can be realized for sub-Planckian excursions for the (non-canonical) inflaton field as well as for sub-Planckian parameter scales.  
Sub-Planckian excursions for the inflaton field are welcome because possible non-renormalizable terms are suppressed. This fact together with the absence of the F-terms  render this model free from the notorious $\eta$-problem of supergravity.  Secondly, the field space region where the slow-roll conditions are satisfied is increased. Hence an inflationary period is realized for more generic initial conditions. Thirdly, the relation between the spectral index of the scalar perturbations and the slow-roll parameters is modified due to the corrected sound speed of the scalar perturbations. 
These imply that some inflationary models may provide a better fit to the data or even render some excluded inflationary models observationally viable.
For example, here, the predictions of the exponential potential can be accommodated by the combined Planck and BICEP2 data \cite{Dalianis:2014nwa}.

Concluding, we comment that throughout this work we have considered single field inflationary potentials. 
Although their inflationary dynamics are influenced by the value of  the new scale $M_*$, these potentials are characterized only by Planck mass suppression scale.

We believe that this work has offered some different insight into how inflation might work in a supergravity framework.

\section*{Acknowledgments}

We thank C. Germani, A. Kehagias, E.N. Saridakis and R. von Unge for comments and discussion.
This work is supported by the Grant agency of the Czech republic under the grant P201/12/G028 and the ''Operational Programme Education and Lifelong Learning'' 
of the European Social Fund and Greek National  Resources.

\end{document}